\documentclass[twocolumn,showpacs,preprintnumbers,amsmath,amssymb]{revtex4-2}
\usepackage{graphicx}
\usepackage{dcolumn}
\usepackage{bm}
\begin{document}
\title{Inequivalence of the low-density insulating state and quantum Hall insulating states in a strongly correlated two-dimensional electron system}
\author{M.~Yu.~Melnikov, D.~G. Smirnov, and A.~A. Shashkin}
\affiliation{Institute of Solid State Physics, Chernogolovka, Moscow District 142432, Russia}
\author{S.-H. Huang and C.~W. Liu}
\affiliation{Department of Electrical Engineering and Graduate Institute of Electronics Engineering, National Taiwan University, Taipei 106, Taiwan}
\author{S.~V. Kravchenko}
\affiliation{Physics Department, Northeastern University, Boston, Massachusetts 02115, USA}
\begin{abstract}
We find that the behaviors of the voltage-current characteristics as one enters the low-density insulating state and integer quantum Hall insulating states in the ultra-clean two-dimensional electron system in SiGe/Si/SiGe quantum wells are qualitatively different. The double-threshold voltage-current curves, representative of the electron solid formation at low densities, are not observed in the quantum Hall regime, which does not confirm the existence of a quasi-particle quantum Hall Wigner solid and indicates that quasi-particles near integer filling do not form an independent subsystem.
\end{abstract}
\maketitle
\setlength{\parskip}{0pt}

The ground state of a two-dimensional (2D) electron system in the strongly interacting limit at low densities is expected to be a Wigner crystal \cite{chaplik1972possible}. It has been predicted that the application of a perpendicular magnetic field should stabilize the crystalline state by suppressing the amplitude of the zero-point vibrations of the electrons at the lattice sites \cite{lozovik1975crystallization,ulinich1978phase,fukuyama1975two,eguiluz1983two}. The Wigner crystal has been predicted to form in a single-valley 2D electron system at Landau filling factors below approximately 0.15 \cite{lam1984liquid,levesque1984crystallization}. There have been claims of the observation of a magnetically-induced Wigner crystal on semiconductor surfaces, which were based on the observation of reentrant insulating phases and single-threshold current-voltage ($I$-$V$) characteristics in the insulating phases around the integer and fractional quantum Hall states, attributed to the depinning of the Wigner crystal (see, \textit{e.g.},
Refs.~\cite{andrei1988observation,williams1991conduction,diorio1992reentrant,qiu2012connecting,knighton2014reentrant,qiu2018new,knighton2018evidence,hossain2022anisotropic,falson2022competing,madathil2023moving}). However, alternative mundane interpretations of the data were discussed, such as Efros-Shklovskii variable range hopping in strong electric fields \cite{marianer1992effective} or percolation \cite{jiang1991magnetotransport,dolgopolov1992metal,shashkin1994insulating}. There have also been claims of the detection of the Wigner crystal at low filling factors in the 2D carrier system in AlGaAs/GaAs heterojunctions, based on the observation of resonance in the real part of frequency-dependent diagonal microwave conductivity that was interpreted as the pinning mode of Wigner crystal domains oscillating in the disorder potential (see, \textit{e.g.}, Refs.~\cite{williams1991conduction,engel1997microwave,ye2002correlation,sambandamurthy2006pinning,moon2014pinning,freeman2024origin}).

Similar resonances were observed in AlGaAs/GaAs 2D electron systems in the quantum Hall regime near integer and fractional Landau fillings and interpreted as originating from a Wigner crystal state formed by quasi-particles with density determined by the deviation from the integer or fractional filling factor \cite{chen2003microwave,lewis2004evidence,lewis2004wigner,zhu2010observation,hatke2014microwave,moon2015microwave,kim2021the}. The formation of a 2D Wigner crystal was assumed in the interpretation of measurements of the changes in the chemical potential with density variation in the integer quantum Hall regime \cite{zhang2014signatures} and measurements of the nuclear magnetic resonance in the integer and fractional quantum Hall regimes \cite{tiemann2014nmr}. The observation of local maxima and minima in the longitudinal resistance near the filling factor $\nu=1$, which is referred to as the reentrant integer quantum Hall states, was interpreted under the assumption that an integer quantum Hall Wigner solid exists \cite{liu2012observation,liu2014fractional,myers2021magnetotransport,myers2024thermal,huang2025density}. Notably, the Wigner crystal states in question, which are expected to spontaneously form without any periodic potential, differ from the generalized Wigner crystal states on a lattice background discussed in Refs.~\cite{li2021imaging,tsui2024direct}.

Recently, two-threshold $V$-$I$ characteristics that reveal the signature of a quantum Wigner solid and exclude mundane interpretations in terms of percolation or overheating have been observed at low densities in 2D electron systems in high-mobility silicon metal-oxide-semiconductor field-effect transistors in a zero magnetic field \cite{brussarski2018transport} and in SiGe/Si/SiGe quantum wells in both zero and perpendicular magnetic fields \cite{melnikov2024collective,melnikov2025stabilization,shashkin2025transport}. The $V$-$I$ characteristics are strikingly similar to those observed for the collective depinning of the vortex lattice in type-II superconductors \cite{blatter1994vortices} with the voltage and current axes interchanged. The results can be described by a phenomenological theory of the collective depinning of elastic structures, which corresponds to thermally activated transport accompanied by a peak in generated broadband current noise between the dynamic and static thresholds. The solid slides with friction as a whole over a pinning barrier above the static threshold.

Assuming that the quasi-particle quantum Hall Wigner solid actually does exist, one expects similar data in the low-density insulating state and quantum Hall insulating states, according to Refs.~\cite{chen2003microwave,lewis2004evidence,lewis2004wigner,zhu2010observation,hatke2014microwave,moon2015microwave,kim2021the}. In this Letter, we find that the behaviors of the voltage-current characteristics as one enters the low-density insulating state and integer quantum Hall insulating states in the ultra-clean two-dimensional electron system in SiGe/Si/SiGe quantum wells are qualitatively different. The double-threshold voltage-current curves, representative of the electron solid formation at low densities, are not observed in the quantum Hall regime, which does not confirm the existence of a quasi-particle quantum Hall Wigner solid. We conclude that the quasi-particles near integer filling do not form an independent subsystem.

Data were obtained on ultra-high-mobility SiGe/Si/SiGe quantum wells similar to those described in Refs.~\cite{melnikov2015ultra,melnikov2017unusual,melnikov2024triple}. The low-temperature electron mobility in these samples reaches approximately 200~m$^2$/Vs at a density about $1\times 10^{11}$~cm$^{-2}$. The approximately 15~nm wide silicon (001) quantum well is sandwiched between Si$_{0.8}$Ge$_{0.2}$ potential barriers. The samples were patterned in Hall-bar shapes with a distance between the potential probes of 100~$\mu$m and a width of 50~$\mu$m using photolithography. The triple top-gate design included the main Hall-bar gate, the contact gate, and the depleting gate, separated by SiO as an insulator; this dramatically reduced the contact resistances and suppressed the shunting channel between the contacts outside the Hall bar, which manifested itself at the lowest electron densities in the insulating regime (for more details, see Ref.~\cite{melnikov2024triple}). No additional doping was used, and the electron density was controlled by applying a positive dc voltage to the gate relative to the contacts. Measurements were performed in an Oxford TLM-400 dilution refrigerator equipped with a magnet. For measurements in the low-density insulating state, the voltage was applied between the Hall potential probes. The current was measured by a current-voltage converter connected to a digital voltmeter. The voltage-current curves displayed slight asymmetry upon reversal of the voltage. We plotted the negative part versus the absolute voltage value for ease of representation. For measurements in the integer quantum Hall insulating states, the current was passed between the source and drain contacts, and the longitudinal voltage was measured by a differential preamplifier connected to a digital voltmeter. The electron density was determined by analyzing Shubnikov-de Haas oscillations in the metallic regime using a standard four-terminal lock-in technique. We applied saturating infrared illumination to the samples for several minutes, after which the quality of the contacts improved and the electron mobility increased, as had been empirically observed \cite{melnikov2015ultra,melnikov2017unusual}. Contact resistances measured below 10 k$\Omega$.

\begin{figure}
\hspace{-1mm}\scalebox{.7}{\includegraphics[width=\columnwidth]{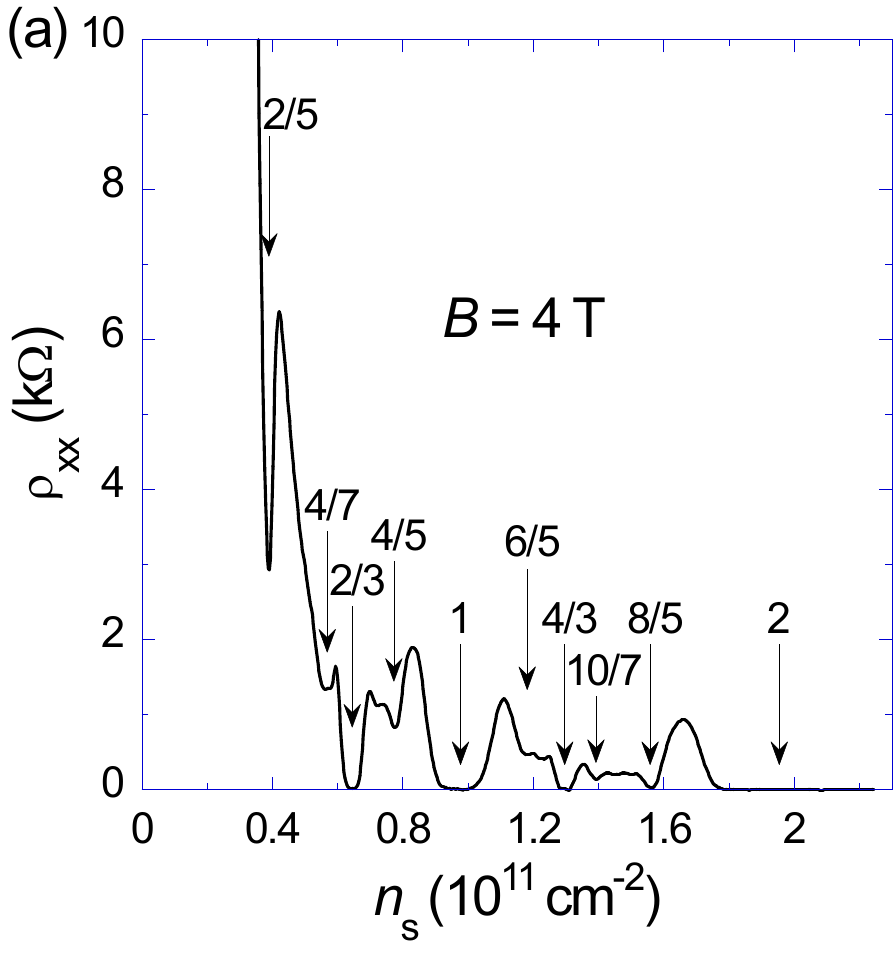}}
\scalebox{.7}{\includegraphics[width=\columnwidth]{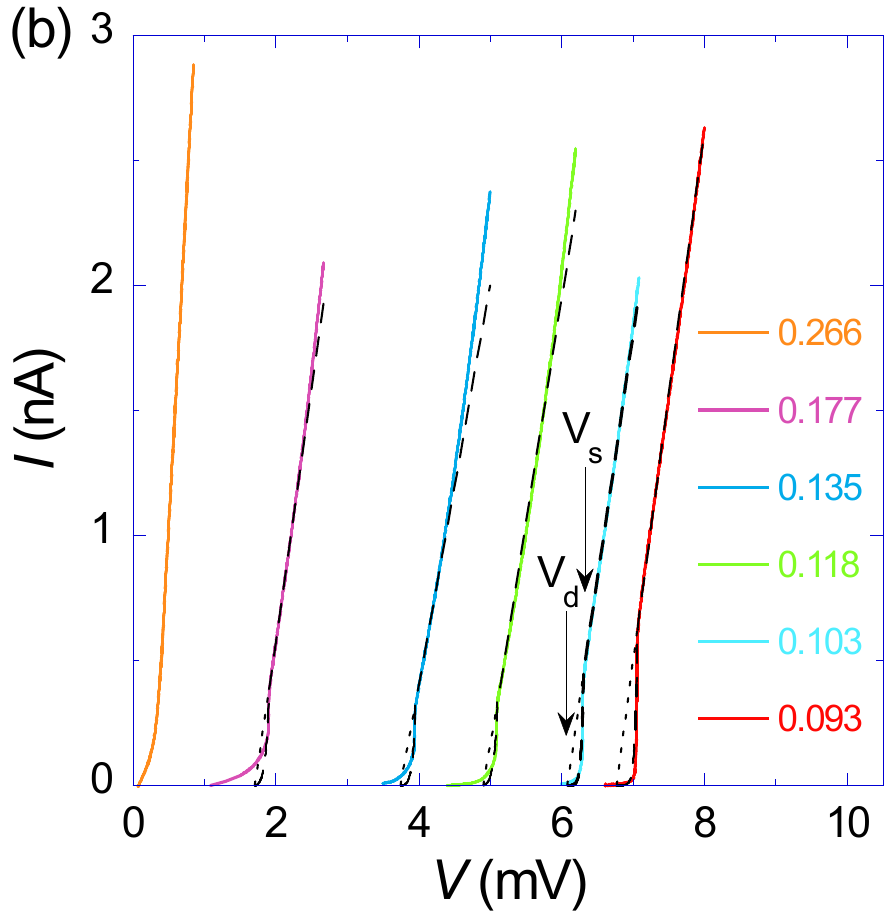}}
\caption{(a) Magnetoresistivity as a function of electron density at $B=4$~T and $T=30$~mK. The minima of the oscillations are indicated. (b) Voltage-current characteristics at different electron densities in the low-density insulating state at $B=4$~T and $T=30$~mK. The electron densities are indicated in units of $10^{11}$~cm$^{-2}$. Also shown are the dynamic threshold $V_{\text d}$ obtained by the extrapolation (dotted line) of the linear part of the $V$-$I$ curves to zero current and the static threshold $V_{\text s}$.  The dashed lines are fits to the data, see text.}
\label{fig1}
\end{figure}

\begin{figure}
\hspace{-1mm}\scalebox{.68}{\includegraphics[width=\columnwidth]{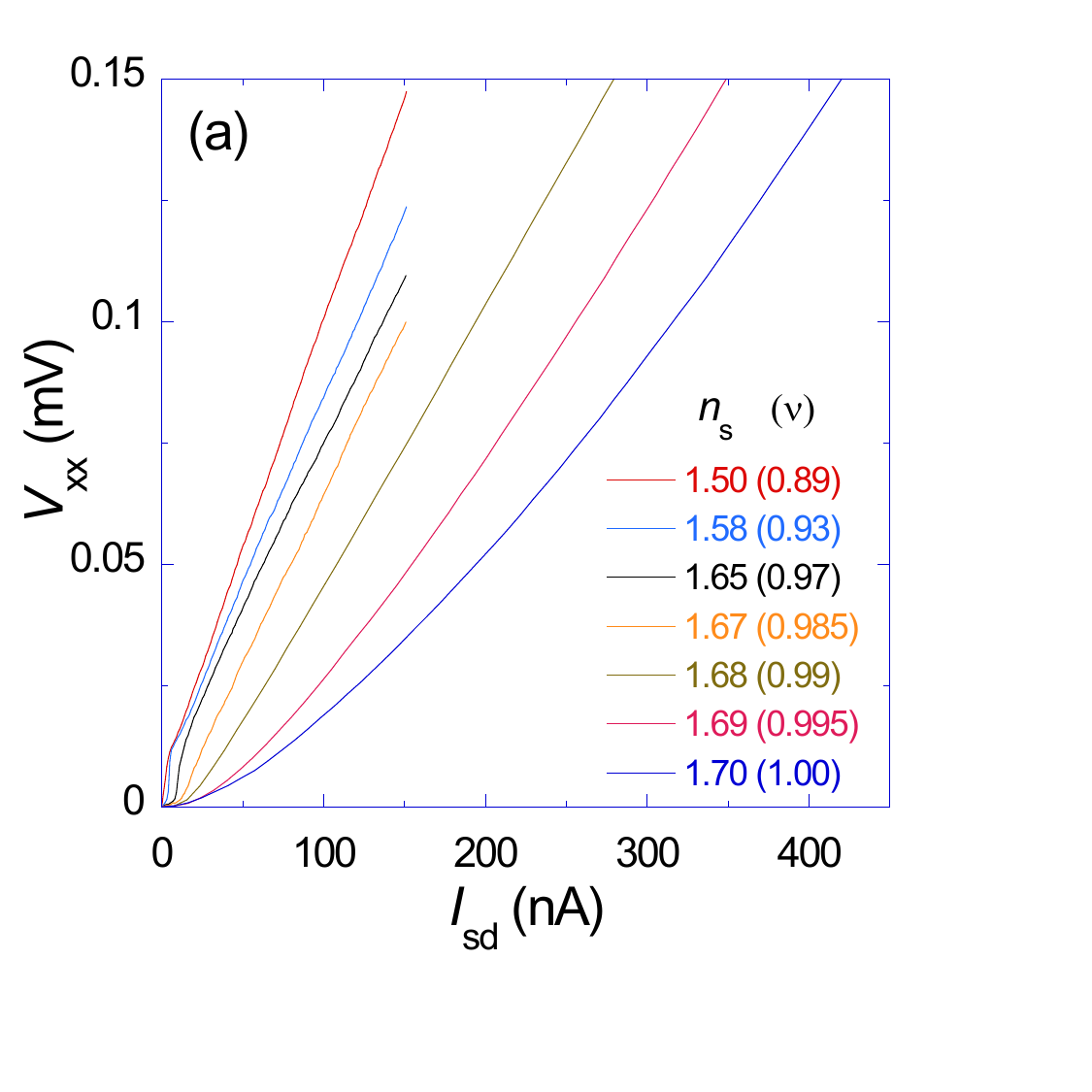}}
\scalebox{.7}{\includegraphics[width=\columnwidth]{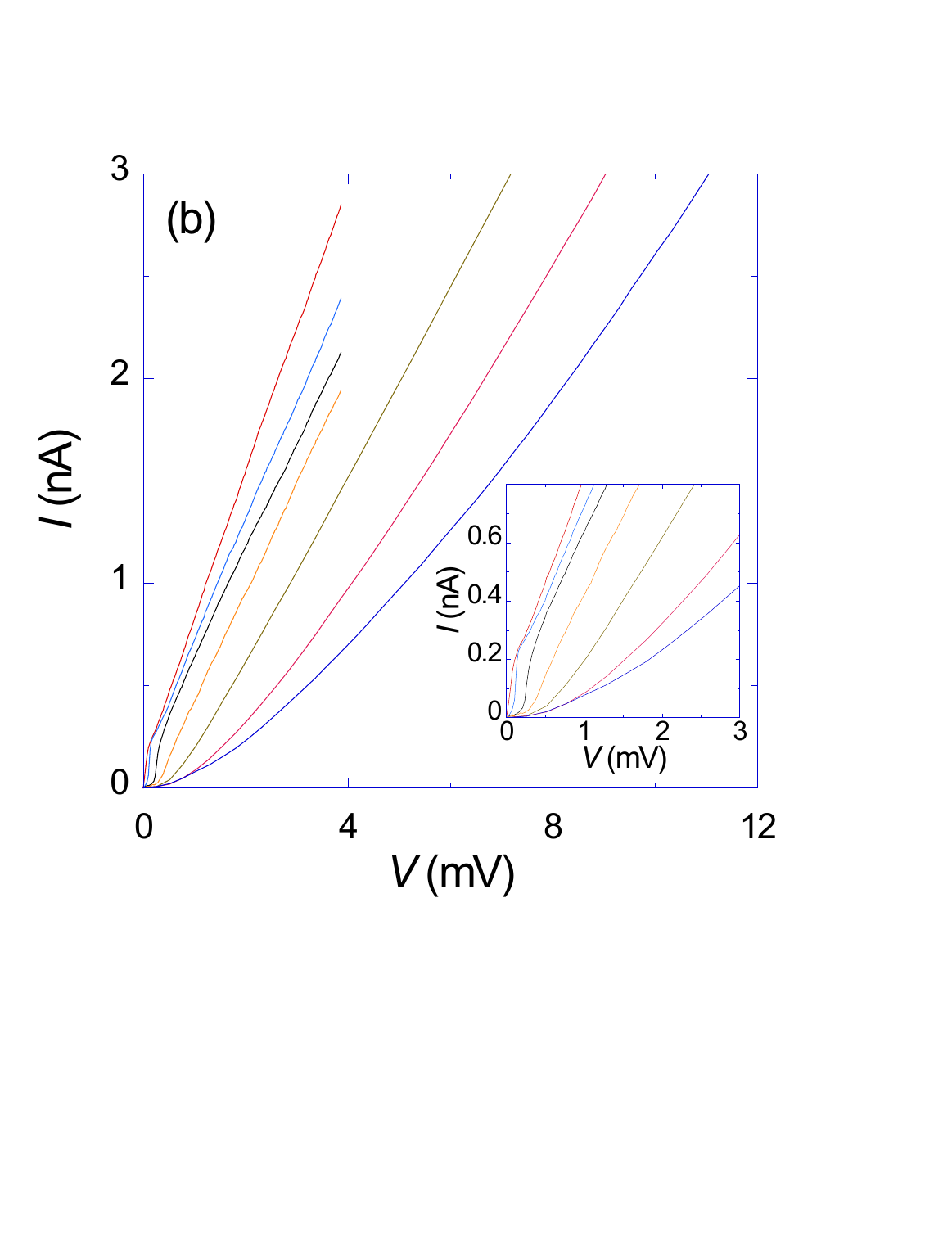}}
\caption{(a) The longitudinal voltage as a function of source-drain current at different electron densities in a magnetic field of $B=7$~T and at a temperature of $T=30$~mK for $\nu\leq 1$. The electron densities are indicated in units of $10^{11}$~cm$^{-2}$, along with the filling factors in brackets. (b) Voltage-current characteristics recalculated from the data shown in (a). The inset shows the data on an expanded scale.}
\label{fig2}
\end{figure}

In Fig.~\ref{fig1}(a), we show the magnetoresistivity $\rho_{xx}$ as a function of electron density $n_s$ measured in the metallic regime in a perpendicular magnetic field $B=4$~T at $T=30$~mK. The pronounced minima of the magnetoresistivity oscillations are observed at filling factor $\nu=n_shc/eB=1$, 2, 2/3, 2/5, 4/5, 4/7, 4/3, 6/5, 8/5, 10/7. This data confirms the high quality of the samples. The voltage-current ($V$-$I$) characteristics measured at different electron densities in the low-density insulating state in $B=4$~T at $T=30$~mK are represented in Fig.~\ref{fig1}(b). Two-threshold voltage-current curves are observed at electron densities below $n_s\approx 0.26\times 10^{11}$~cm$^{-2}$. As the applied voltage increases, the current remains near zero until the first threshold voltage is reached. Then, the current increases sharply until the second threshold voltage is reached, above which the slope of the voltage-current curves is significantly reduced, and the behavior becomes linear, although not ohmic, consistent with the results obtained previously \cite{brussarski2018transport,melnikov2024collective,melnikov2025stabilization,shashkin2025transport}. The double-threshold $V$-$I$ curves shift to higher voltages as one advances into the low-density insulating state by reducing the electron density.

\begin{figure}
\hspace{-1mm}\scalebox{.74}{\includegraphics[width=\columnwidth]{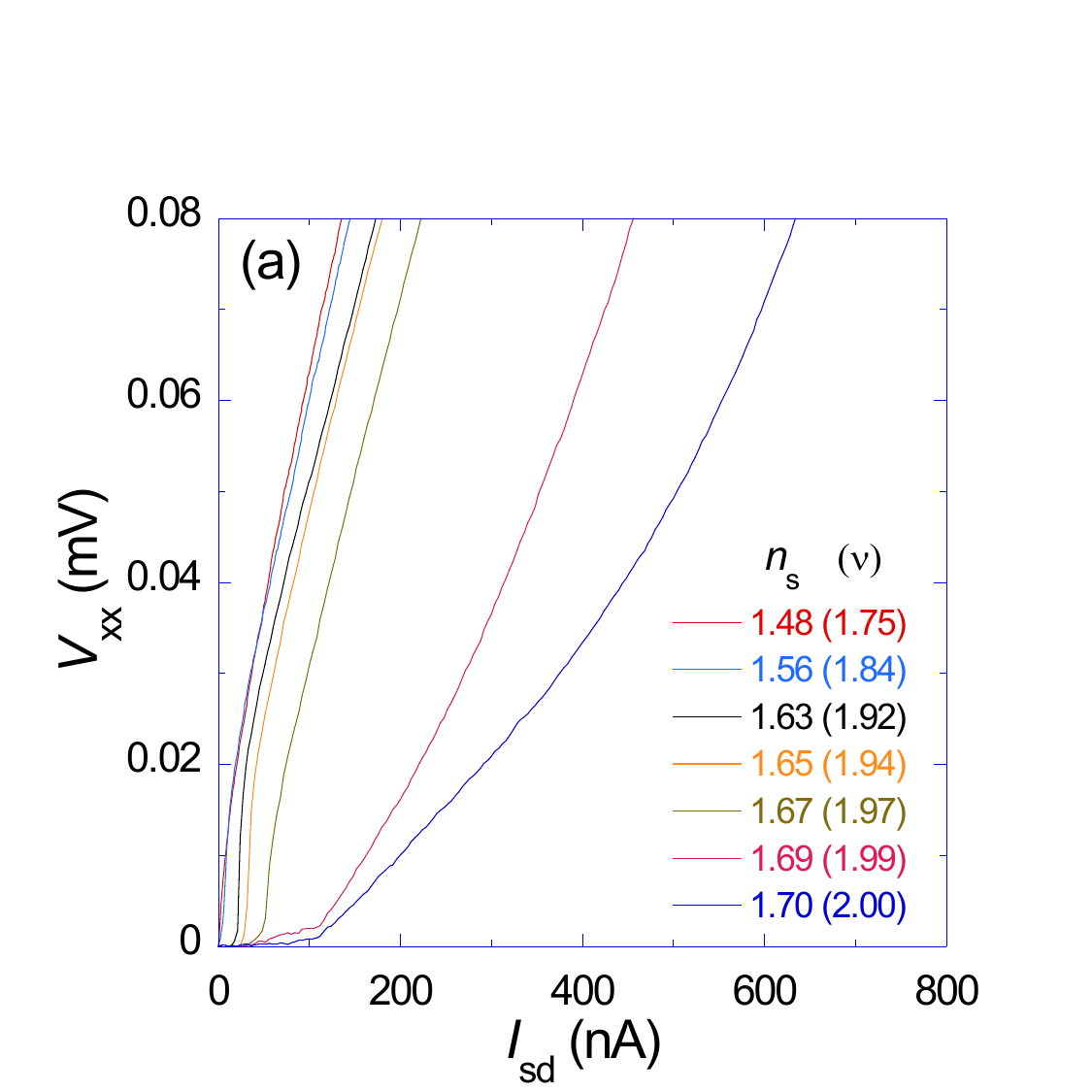}}
\scalebox{.73}{\includegraphics[width=\columnwidth]{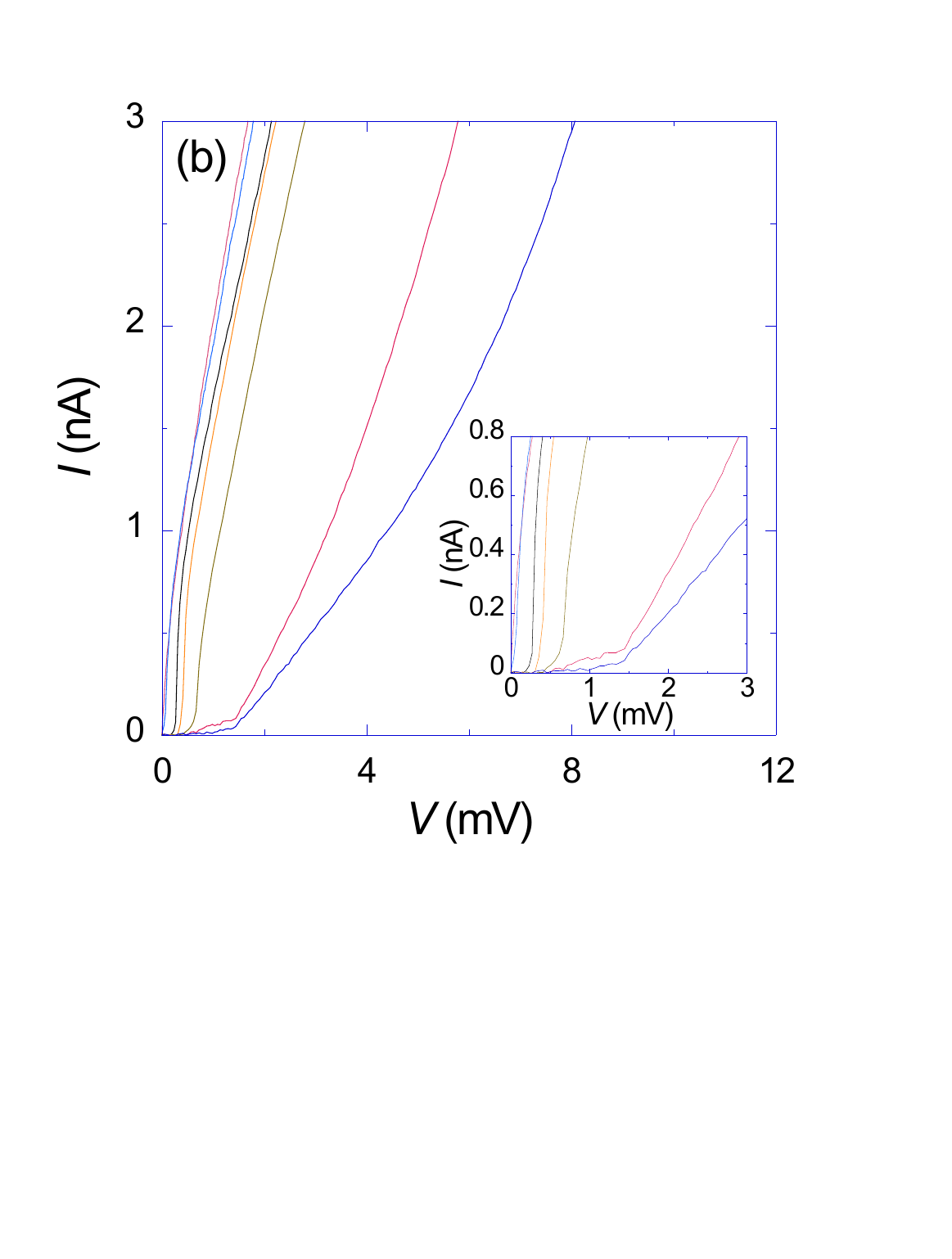}}
\caption{(a) The longitudinal voltage as a function of source-drain current at different electron densities in a magnetic field of $B=3.5$~T and at a temperature of $T=30$~mK for $\nu\leq 2$. The electron densities are indicated in units of $10^{11}$~cm$^{-2}$, along with the filling factors in brackets. (b) Voltage-current characteristics recalculated from the data shown in (a). The inset shows the data on an expanded scale.}
\label{fig3}
\end{figure}

A phenomenological theory of the collective depinning of elastic structures was adapted for an electron solid in Ref.~\cite{brussarski2018transport}. As the applied voltage increases, the depinning of the electron solid is indicated by the appearance of a current. Between the dynamic ($V_{\text d}$) and static ($V_{\text s}$) thresholds, the collective pinning of the electron solid occurs, and the transport is thermally activated: $I=\sigma_0\,(V-V_{\rm d})\,\exp[-U_{\rm c}(1-V/V_{\rm s})/k_{\rm B}T]$, where $U_{\rm c}$ is the maximal activation energy of the pinning centers, $\sigma_0$ is a coefficient, and $V_{\rm d}$ corresponds to the pinning force. When the voltage exceeds the static threshold, the electron solid slides with friction over a pinning barrier, as determined by the balance of the electric, pinning, and friction forces, resulting in linear voltage-current characteristics: $I=\sigma_0\,(V-V_{\rm d})$. The corresponding fits, shown by the dashed lines in Fig.~\ref{fig1}(b), accurately describe the experimental two-threshold $V$-$I$ characteristics.

In Figs.~\ref{fig2}(a) and \ref{fig3}(a), we show the breakdown dependences of the longitudinal voltage $V_{xx}$ on source-drain current $I_{sd}$ at different electron densities in a fixed magnetic field at $T=30$~mK for the quantum Hall insulating states near $\nu=1$ and $\nu=2$. For easier presentation, the data are plotted at the electron densities corresponding to filling factors $\nu$ below the integer $\nu_0$; the data obtained at $\nu>\nu_0$ are similar due to symmetry about the integer $\nu_0$. It is easy to recalculate the breakdown curves $V_{xx}(I_{sd})$ in the voltage-current characteristics $I(V)$, as long as the magnetoresistivity $\rho_{xx}$ is much smaller than the Hall resistivity $\rho_{xy}=h/\nu_0 e^2$, where the values refer also to the nonlinear regime. The dissipative current $I$ that flows between the opposite edges of the sample is balanced by the Hall current in the filled Landau levels associated with the longitudinal voltage $V_{xx}$. The quantized value of $\rho_{xy}$ is a factor that allows determination of $I=V_{xx}/\rho_{xy}$ and the Hall voltage $V=I_{sd}\rho_{xy}$ from the experimental breakdown dependence $V_{xx}(I_{sd})$; note that the measured value $V_{xx}$ should be normalized by the aspect ratio equal to 2 for our case. The dependence $I(V)$ is a voltage-current characteristic equivalent to the Corbino geometry case \cite{shashkin1994insulating}. The so-determined $V$-$I$ curves are shown in Figs.~\ref{fig2}(b) and \ref{fig3}(b). These characteristics look similar near $\nu=1$ and $\nu=2$. At the maximum deviations $\Delta n_s=|\nu-\nu_0|eB/hc$, the current increases sharply with increasing applied voltage, and then the slope of the $V$-$I$ curves decreases, corresponding to an approximately proportional increase of the current with voltage. As one enters the quantum Hall insulating state by decreasing the deviation $\Delta n_s$, the initial steep rise of the current and the reduction in the slope of the $V$-$I$ curves disappear. The obtained $V$-$I$ curves are in contrast to the double-threshold $V$-$I$ characteristics in the low-density insulating state from Fig.~\ref{fig1}(b). We emphasize that the qualitative difference between the $V$-$I$ curves in the low-density insulating state and quantum Hall insulating states is observed in similar ranges of voltages, currents, and electron densities ($n_s$) or quasi-particle densities ($\Delta n_s$).

Assuming the existence of a Wigner crystal state formed by quasi-particles with density determined by the deviation from the integer filling factor, one expects similar data in the low-density insulating state and quantum Hall insulating states, according to Refs.~\cite{chen2003microwave,lewis2004evidence,lewis2004wigner,zhu2010observation,hatke2014microwave,moon2015microwave,kim2021the}. We emphasize that the key idea underlying this assumption is that the quasi-particles can be considered separately as an independent subsystem. However, the experimental results obtained in the 2D electron system in SiGe/Si/SiGe quantum wells do not confirm the occurrence of a quasi-particle quantum Hall Wigner solid. The low-density insulating state is characterized by the observed double-threshold voltage-current curves, which serve as a signature of the quantum Wigner solid \cite{brussarski2018transport,melnikov2024collective,melnikov2025stabilization,shashkin2025transport}. In contrast, significantly different $V$-$I$ curves are observed in the integer quantum Hall insulating states. This finding indicates that the quasi-particles near integer filling do not form an independent subsystem. The possible reason may be the mixing of Landau levels \cite{ando1982electronic} due to electron-electron interactions that strongly exceed the cyclotron energy in this 2D electron system. For comparison, the 2D electron system in AlGaAs/GaAs heterostructures, where similar data were reported in the low-density insulating state and quantum Hall insulating states, is characterized by electron-electron interactions that are comparable to the cyclotron energy due to an appreciably smaller (by a factor of about 3) effective mass, in which case the mixing of Landau levels is expected to be less relevant. This difference can be essential in interpreting the experimental results in both electron systems.

In summary, we have found that the behaviors of the voltage-current characteristics as one enters the low-density insulating state and integer quantum Hall insulating states in the ultra-clean two-dimensional electron system in SiGe/Si/SiGe quantum wells are qualitatively different. The double-threshold voltage-current curves, representative of the electron solid formation at low densities, are not observed in the quantum Hall regime, which does not confirm the existence of a quasi-particle quantum Hall Wigner solid. We conclude that the quasi-particles near integer filling do not form an independent subsystem. Certainly, further experiments are needed on other 2D electron systems.

We gratefully acknowledge discussions with K.-S. Kim, S.~A. Kivelson, and B. Spivak. The ISSP group was supported by the RF State Task. The NTU group acknowledges support by the Ministry of Science and Technology, Taiwan (Project No.\ NSTC 113-2634-F-A49-008). S.V.K. was supported by NSF Grant No.\ 1904024.


%

\end{document}